\documentclass[twocolumn,showpacs,preprintnumbers,superscriptaddress]{revtex4}

\usepackage{graphicx}
\usepackage{dcolumn}
\usepackage{bm}

\begin{document}

\preprint{Chaos {\bf 13}, 319 (2003)}

\title{Experimental Study of Imperfect Phase Synchronization in the Forced
Lorenz System}

\author{Antonio Pujol-Per\'e}
\homepage{http://www.imedea.uib.es/PhysDept/}
\affiliation{Instituto Mediterr\'aneo de Estudios Avanzados, IMEDEA (CSIC-UIB),
E-07071 Palma de Mallorca, Spain}

\author{Oscar Calvo}
\email{oscar@galiota.uib.es}
\affiliation{Departament de F\'{\i}sica, Universitat de les Illes Balears,
E-07071 Palma de Mallorca, Spain}

\author{Manuel A.\ Mat\'{\i}as}
\email{manuel@imedea.uib.es}
\affiliation{Instituto Mediterr\'aneo de Estudios Avanzados, IMEDEA (CSIC-UIB),
E-07071 Palma de Mallorca, Spain}

\author{J\"urgen Kurths}
\email{juergen@agnld.uni-potsdam.de}
\affiliation{Institute of Physics, Potsdam University, PF 601553, D-14115 Potsdam, 
Germany}

\date{\today}

\begin{abstract}
Recent studies have illustrated a phenomenon that occurs in certain sinusoidally forced
chaotic oscillators: (chaotic) phase synchronization, in which the two, quite
different systems, oscillate at the same pace. Imperfect phase synchronization appears
in oscillators exhibiting unbounded return times, {\it e.g.}, oscillators
in which the chaotic set includes a saddle equilibrium, as is the case of Lorenz
oscillator. We demonstrate the phenomenon of imperfect phase synchronization in an
experimental system: an analog Lorenz circuit, including its implications in the
behavior of the system.
\end{abstract}


\maketitle


{\bf 
A class of (so-called phase coherent) chaotic oscillators, namely R\"ossler oscillator,
has been shown to exhibit phase synchronization in the case that the oscillator is driven
by a sinusoidal generator (and also in the case of two, slightly detuned, chaotic oscillators).
This behavior is characterized by an approximately constant relationship between a
suitably defined phase for the chaotic oscillator and the phase of the sinusoidal
generator. Interestingly, the oscillator remains chaotic, and so does the amplitude,
while its rhythm is dictated by the external sinusoidal generator, and, thus, is much
more regular. Quite different is the case of chaotic 
oscillators for which a saddle equilibrium belongs to the attractor, as is the case of 
Lorenz oscillator. The most relevant feature of this type of systems
is that a typical trajectory in phase space has some probability of passing close
enough to the stable manifold of the saddle point (in the Lorenz system this happens
whenever a trajectory changes lobe). The closer the trajectory approaches
the stable manifold of the saddle point, the longer is the return time, {\it i.e.}, the
time needed to perform a turn. Ultimately, these extra long return times (compared to the
{\it typical\/} return times of the system, and also to the period of the external
sinusoidal generator) make it difficult to achieve the state of (perfect) phase synchronization,
leading to the behavior known as imperfect phase synchronization. Here we shall demonstrate how
this behavior is typical, in the sense that it can be easily reproduced in a experimental
implementation of the Lorenz oscillator. 
}

\section{Introduction} \label{secintrod}

Recently, there has been a lot of interest in the study of manifestations of
synchronization in several physical, chemical, biological, and technological
systems \cite{syncbook}. Probably the simplest (and most studied) situation corresponds
to a (dynamical) system forced by a sinusoidal generator. In this context, synchronization 
is understood as the readjustment in the rhythm of the forced system under the influence
of the driving signal. In the periodic case this was already studied by Arnold, and 
then by many others (see, {\it e.g.}, \cite{ottbook}), and the main features of this behavior have
been uncovered. In particular, as the coupling becomes different from zero one expects regions of
parameters for which synchronization (or phase locking) occurs. If one represents
the amplitude versus the frequency, both corresponding to the sinusoidal forcing,
one obtains the well-known Arnold tongues, namely wedge-like regions of synchronized
behavior. 

The situation is somehow more complex if one considers systems with chaotic behavior.
For relatively strong coupling, it was already shown \cite{fujisaka,pecora90} that
one may have complete synchronization between identical, uni- or bi-directionally
coupled chaotic oscillators. Generalized synchronization \cite{Rulkov95}, implying
a functional relationship between drive and response has been also found for
uni-directionally coupled chaotic systems. More recently a type of partial
synchronization was shown for bi-directionally (slightly detuned) coupled oscillators: 
phase synchronization \cite{rosenblum96}. The chaotic systems that have been shown to
exhibit this behavior ({\it e.g.}, R\"ossler system \cite{rossler}) can be considered as true
oscillators,  in the sense that the systems exhibit oscillations in
phase space (around some center of oscillation). This implies also that a phase
variable can be suitably defined \cite{yalcin97,pikovsky97a}, and the observed
behavior is that there is some regime in which the two systems share the
phase (apart from a constant, smaller than $2\pi$), while the amplitudes
vary chaotically and are practically uncorrelated
\cite{rosenblum96}. The concept of phase synchronization has shown
to be useful, although it cannot be applied in general for an arbitrary dynamical system,
and, in particular, allows to study synchronization behaviors where not much
information can be obtained by looking at correlations between the coupled
systems. 

Phase synchronization has also been found in the case of sinusoidally forced chaotic
oscillators, and this will be the focus of our study. The chaotic
oscillator and the drive are not homologous, but it can be shown that one may 
get phase synchronization \cite{pikovsky97a}, in the sense
that a suitably defined phase for the chaotic oscillator minus the phase of
the drive are bounded by $2\pi$. Chaotic systems with low phase diffusion \cite{josicmar}
({\it e.g.}, the R\"ossler system) exhibit, in principle, perfect phase synchronization
behavior under sinusoidal forcing. Quite different is the situation if one works
with the Lorenz \cite{lorenz} system (at the parameter values for which it exhibits the
well known butterfly attractor).
For the butterfly Lorenz system the saddle equilibrium point at the origin is part of
the closure of the attractor, and makes the attractor non-hyperbolic by inducing
singularities for the return maps. In particular, the return times to a suitably
defined Poincar\'e cross section will exhibit a singularity, corresponding to the
crossing of the return map with the stable manifold of the saddle equilibrium, that
happens sometimes when the Lorenz system changes lobe. 


Thus, it is normal to expect that phase synchronization will not be {\it perfect\/}
for a driven Lorenz system, in the sense, that the system will not be able to
follow the pace of the drive at all time, namely when passing close to the saddle
equilibrium. This has been, indeed, recently
shown through numerical simulation and theoretical arguments by Zaks {\it et al.}
\cite{zaks99,park99,zaks00}. This {\it imperfect\/} phase synchronization
manifests, among other effects, in the presence of phase slips, that are jumps
by $2\pi$ in the phase. It must be pointed out that phase jumps may also be obtained 
in, at least, two other different circumstances, namely in the presence of noise,
and for parameter values close to the onset of phase synchronization. In the
first case the stochastic (high-dimensional) degrees of freedom may induce
occasional kicks out of the {\it synchronized state\/} exhibiting some kind of
higher-dimensional behavior, while in the second the phase locked stable and
unstable solutions, respectively, will collide leading to a so-called {\it eyelet}
intermittency \cite{Pikovsky97b}. Instead, imperfect phase
synchronization is a behavior in which a purely deterministic system exhibits
a non-stationary behavior, not associated to external influences or the proximity
to the onset of phase synchronization. 

An alternative way of understanding phase synchronization is in terms of 
unstable periodic orbits (UPOs) \cite{Pikovsky97d}. In the case of phase coherent 
systems (systems with a relatively narrow distribution of return times, {\it i.e.}, 
of frequencies), phase synchronization is attained when all the UPOs become entrained 
with the forcing (around the {\it natural\/} frequency of the system), and this
is possible for all the UPOs simultaneously as they have similar frequencies.
In the case of the Lorenz butterfly system (and in general systems with a broad distribution 
of return times), and due to the influence of the saddle equilibrium point at the origin,
it is not possible to find conditions in which all the UPOs become simultaneously
entrained with the forcing (even for the natural frequency) at a fixed, established
locking ratio ({\it e.g.}, 1:1), but epochs of synchronized behavior (sometimes 
long) are interspersed with periods of time for which remains out of sync.

One of the findings of Refs. \cite{zaks99,park99} is that the system actually exhibits
synchronization at almost all time, but with time epochs characterized by
different (alternating) locking ratios (that correspond to the number of turns of the chaotic
oscillator with respect to the sinusoidal oscillator).
Thus, phase jumps for imperfect phase synchronization exhibit 
distinctive features when compared to phase jumps due to noise or eyelet intermittent behavior.
This property of imperfect phase synchronization is very important when considering
an experimental system (as is our case) subject to many sources of unavoidable
experimental noise, like thermal noise, channel noise, etc. In this sense, we will
show that the observed phase jumps have a clear deterministic structure,
corresponding to alternate locking ratios, quite different to the effect of external
noise or proximity to the onset of the transition to phase synchronization.

Another point of interest in our study concerns the ability to model deterministic
chaotic systems as it has been found that in some circumstances \cite{laigrekur99,laigreb99}
these systems may exhibit obstructions to deterministic modeling. Thus, in Ref. 
\cite{laigreb99}
the authors state that {\it $\ldots$ in laboratory experiments ($\ldots$) it
might only make sense to work directly with measured time series instead of a
mathematical model when attempting to understand the long-term behavior of the
system}. These difficulties are a manifestation of nonhyperbolicity, and, from
the reasoning above, they cannot be completely excluded in our system, namely
when a system trajectory approaches the saddle equilibrium. In this sense,
studying the phenomenon of imperfect phase synchronization in a real physical
system is the only way of proving unambiguously its existence.

The goal of this paper is to present the first experimental study of imperfect
phase synchronization for a circuit, that represents the Lorenz system
subject to sinusoidal forcing. Section \ref{secexp} discusses the Lorenz circuit
and the experimental methodology. Section \ref{secres} discusses the main results
of this work, and their comparison with the theoretical study. And, finally,
Section \ref{secconc} contains the main conclusions of the present work.

\section{Experimental system and Method} \label{secexp}

The analog circuit representing the Lorenz system \cite{lorenz} is the one 
described in Ref.\ \cite{sanchezphd,sanchezpre98}. Starting with
the differential equations representing the Lorenz system plus a 
sinusoidal forcing term in the $\dot{z}$ term \cite{zaks99,park99} 
(forcing is introduced in this term in order to preserve the symmetry of
the equations),
\begin{eqnarray}
\begin{array}{rcl}
\dot{x}&=&\sigma(y-x)\\
\dot{y}&=&R\,x-y-x\,z\\
\dot{z}&=&x\,y-b\,\,z+E^{\prime}\,\sin(\Omega^{\prime} t)
\end{array}
\label{eqlor}\ .
\end{eqnarray}
The circuit consists of three integrators, one for each variable, and the
nonlinear terms are represented using analog multipliers. The first step
in designing the circuit is to rescale both the three state variables $x$, $y$,
and $z$ in order to fit within the dynamical range of the source $[-15\,V,
15\,V]$, and such that the circuit operates in the frequency range of a few 
kilohertz. The transformation applied to the variables is the following:
\begin{equation}
u=x/5\quad v=y/5\quad  w=z/10\quad \tau =t/A\quad A=10^3\ 
\label{scaling}
\end{equation}
This rescaling of variables leads to the following set of differential equations,
in which the variables, $u$, $v$, $w$, are voltages across the three
capacitors of the circuit, and in which the time is expressed in seconds,
\begin{eqnarray}
\begin{array}{rcl}
\dot{u}&=&A\,\sigma(v-u)\\
\dot{v}&=&A\,(R\,u-v-10\,u\,w)\\
\dot{w}&=&A\,[(2.5\,u\,v-b\,\,w)+E \sin(\Omega\,\tau)]
\end{array}
\label{eqlorsc}\ .
\end{eqnarray}
where $E=E^{\prime}/10$ and $\Omega=A\,\Omega^{\prime}=10^3\,\Omega^{\prime}$. 
In Eq.\ (\ref{eqlorsc}) $\dot{u}=du/d\tau$, $\dot{v}=dv/d\tau$, and 
$\dot{w}=dw/d\tau$, as the time is expressed in rescaled units. 

\begin{figure}
\includegraphics[width=3.4in]{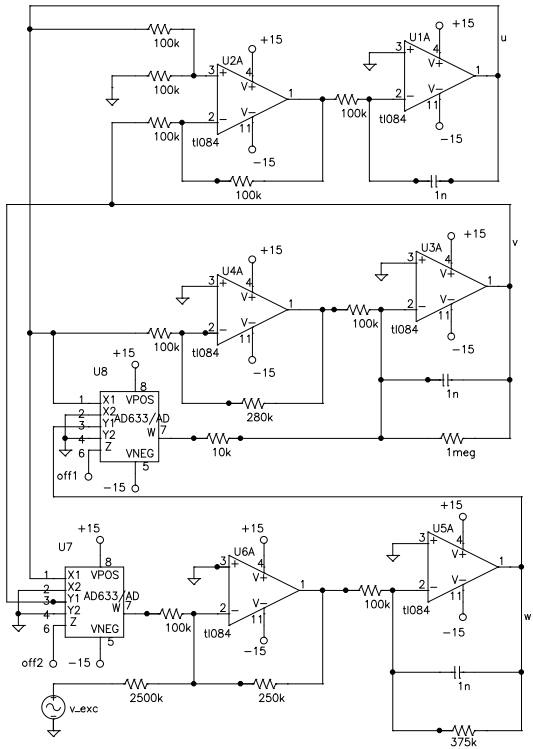}
\caption[]{Schematic representation of the circuit representing the Lorenz oscillator
in rescaled variables, Eq.\ (\ref{eqlorsc}), including the sinusoidal forcing term.}
\label{scheme}
\end{figure}

These equations have been implemented in an electronic circuit as shown in Fig.\
\ref{scheme}. The analog multipliers (AD633) have a noticeable offset at the
output that may alter the dynamical
behavior of the system, and this has been compensated using a compensation
array. The tolerances of the resistors and capacitors are of $1\,\%$
or less. In particular, the parameters for the
Lorenz oscillator (\ref{eqlorsc}) recalculated from the actual values of
the electronic components are as follows: $\sigma=10.19$, $b=2.664$ and $R=28.17$
(to be compared with the intended values: $\sigma=10$, $b=8/3$ and $R=28$).
All the experimental results have been measured with a sampling rate of $80\,{\rm kHz}$
using a data acquisition card with $12\,{\rm bits}$ of resolution, sufficient for the dynamicç
range of the Lorenz
circuit. In all the studies presented here the amplitude of the forcing 
has been fixed (through a resistance) to be $E=1\,V$ (corresponding to $E^{\prime}=10$ 
for the Lorenz system (\ref{eqlor}) before the rescaling). Another important information
concerning the system is the natural frequency of the unforced Lorenz system,
that has been found to be $\omega_0=1311\, Hz=8241\,{\rm rad/s}$. It has been estimated
by using Eq. (2) in Ref.\ \cite{park99}. The above quoted values of $\sigma$, $b$, $R$, and 
$E$ have been kept fixed in all  the results presented in this paper.

\section{Results} \label{secres}

As already mentioned in the introduction, the key feature of the Lorenz system
for the parameter values considered in the present work is that the saddle
point at the origin, $u=v=w=0$ is part of the attractor. This single point
is determinant in the dynamics of the system due to the fact that the dynamics of 
the Lorenz system for the parameter values studied in the present work consists
basically in spiraling around one lobe followed by jumping to the other lobe,
where the system exhibits the same spiraling dynamics, and jumping again. While
the system is rotating in a given lobe these rotations are quite regular (and
{\it fast\/}). Instead, jumping to the other (symmetric) lobe implies that the system
becomes under the influence of the stable manifold of the saddle point at the 
origin, what leads to a slow down in the dynamics.

\begin{figure}
\includegraphics[width=3.2in]{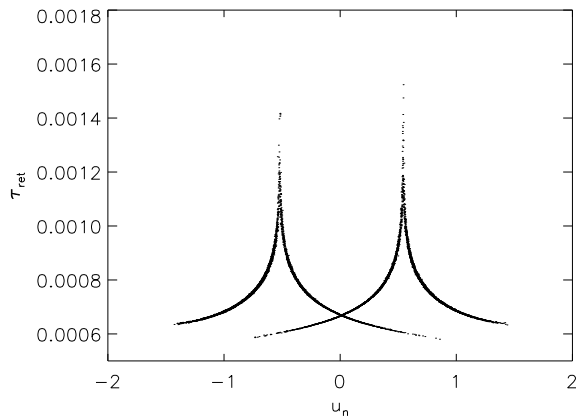}
\caption[]{Return time of the free running Lorenz oscillator at the Poincar\'e
surface $w=(R-1)/10$ versus variable $u$. The two branches correspond 
to the two lobes of the attractor.}
\label{rettime}
\end{figure}

This behavior can be adequately characterized by taking a
suitable Poincar\'e plane $10\,w=z=R-1$, or $w=(R-1)/10$. The (high) rate of
contraction along the transverse direction will lead to an approximately
one-dimensional dynamics in this Poincar\'e section. An interesting characterization
of this behavior can be obtained by representing  the return times at the Poincar\'e
cross section, {\it i.e.}, the times that a trajectory spends between crosses
with the Poincar\'e cross section. As explained, these times are not bounded
from above, and this can be also seen from Fig.\ \ref{rettime}, in which the
time necessary to arrive to the Poincar\'e cross section is represented versus
the value of variable (voltage) $u$ at the crossing. From this
representation it can be clearly seen that the return times diverge logarithmically
when approaching the singularity. 

\begin{figure}
\includegraphics[width=3.2in]{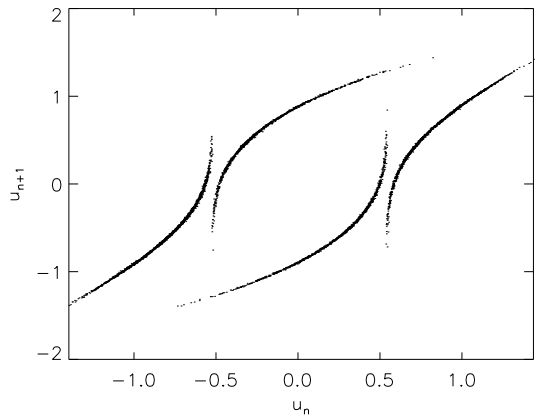}
\caption[]{Return map for the free running Lorenz oscillator at the Poincar\'e
surface $w=(R-1)/10$. Variable $u$ at a given intersection
with the Poincar\'e surface, $u_{n+1}$, is plotted versus the same variable at the
previous intersection, $u_n$. The two parts of the figure correspond to the
two lobes of the attractor.}
\label{retmap}
\end{figure}

As the dynamics at the Poincar\'e cross-section is approximately one-dimensional,
one could consider also a description based on iterated maps, namely by plotting
variable $u$ at a crossing with the Poincar\'e section versus $u$ at the preceding
cross section (see Fig.\ \ref{retmap}). This representation will also exhibit a singularity,
namely at the intersection of the Poincar\'e cross-section with the stable
manifold of the saddle equilibrium.

\begin{figure}
\includegraphics[width=3.2in]{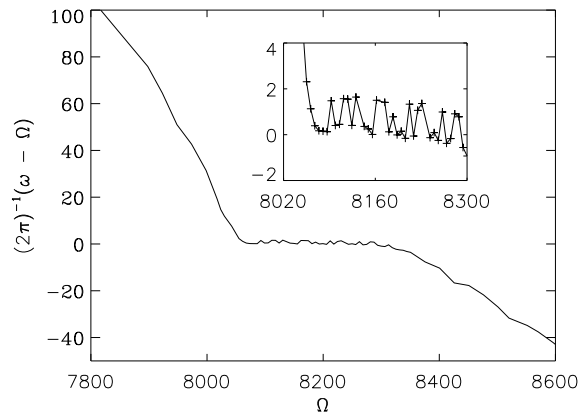}
\caption[]{Difference between mean frequency $\omega$ and driving frequency
$\Omega=$ (in ${\rm kHz}$, estimated from a time of $200\,s$ for each value of the
frequency, corresponding, approximately, to $2.6\times 10^{5}$ turns of the chaotic
oscillator).
}
\label{diffreq}
\end{figure}

As explained above the system studied in this work consists of an oscillator, that due to
its chaotic dynamics exhibits a strong variation in the rotation period, forced
by an oscillator rotating at a fixed pace. The most interesting dynamics of
this system corresponds to those parameter values for which the system
exhibits some kind of synchronization between these two different behaviors.
The type of synchronization found can never be complete
(due to the dissimilar nature of the systems involved), and it is rather phase
synchronization. Thus, both types of oscillations (chaotic and regular)
are different in detail, but beat at the same pace, what implies that they exhibit
approximately the same frequency (this frequency is the average frequency in the case
of the chaotic oscillator). This can be seen from Fig.\ \ref{diffreq}, where the
difference between
the mean frequency of the Lorenz oscillator and the driving frequency is represented.
For a fixed value of the forcing amplitude, $E=1\,V$, and by varying the forcing 
frequency $\Omega$, a region in which the difference of frequencies is quite small 
(close to zero) can be found (cf.\ Fig.\ \ref{diffreq}). A closer inspection (see the
inset of Figure \ref{diffreq}) shows that the plateau is not exactly zero. The
oscillations in the inset (compared to Fig.\ 11 in Ref.\ \cite{park99} should be
ascribed to the larger number of turns used in the latter study, and also to
experimental uncertainties). Anyhow, the frequency difference tends to be positive
in all the synchronization range, as it should (cf. with Fig.\ 11 in Ref.\ \cite{park99}).

\begin{figure}
\includegraphics[width=3.2in]{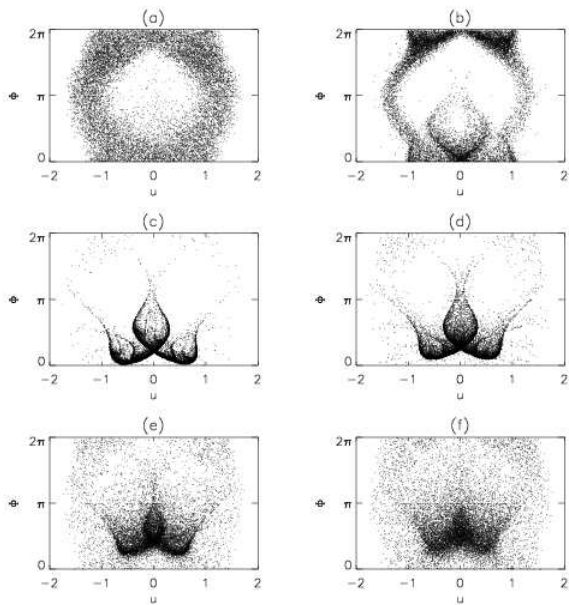}
\caption[]{Snapshot attractors of the Poincar\'e mapping for different values of
$\Omega$: (a) $7350\, {\rm Hz}$; (b) $7900\, {\rm Hz}$; (c) $8150\, {\rm Hz}$;
(d) $8250\, {\rm Hz}$; (e) $8350\, {\rm Hz}$; (f) $8400\, {\rm Hz}$. The phase
$\Phi$ of the sinusoidal oscillator at the Poincar\'e cross section $w=(R-1)/10$
is represented versus variable $u$.
}
\label{poinfig}
\end{figure}

Another quite interesting way of characterizing the imperfect phase synchronization
behavior exhibited by our electronic sinusoidally excited Lorenz oscillator is by
looking at the attractor stroboscopically sampled at a suitable chosen Poincar\'e
section (the result will be a snapshot attractor). In our case we consider
the {\it usual\/} Poincar\'e section $z=R-1$, that in rescaled units becomes
$w=(R-1)/10$, as explained above. The evolution of this snapshot attractor as the
forcing frequency is varied can be seen in Figure \ref{poinfig}. The snapshot attractor
exhibits a transformation from a diffuse cloud for frequencies of the sinusoidal
oscillator outside the synchronization plateau of Figure \ref{diffreq} to a well 
defined pattern inside this synchronization plateau, and, again, a diffuse
cloud when increasing the forcing frequency outside the plateau (see Figure \ref{poinfig}(a--f)).
However, (cf. also Ref.\ \cite{park99}) even inside the {\it well synchronized\/} region 
the snapshot attractor never resembles a (more or less narrow) stripe as expected
for the case of perfect phase synchronization ({\it e.g.}, for the case of a phase
coherent oscillator). As explained in Ref.\ \cite{park99} the well defined pattern
obtained inside the synchronization plateau can be explained noticing that
the system appears to spend most of the time in the central region
(the Figure is symmetric through the change $x\rightarrow -x$ due to the two
lobes exhibited by the attractor), with occasional excursions that form the
{\it whiskers\/} of the pattern.

\begin{figure}
\includegraphics[width=3.2in]{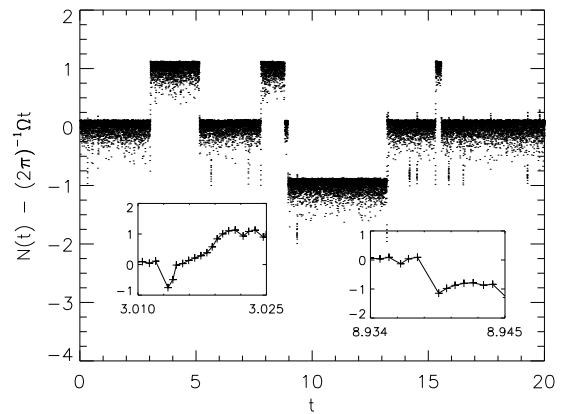}
\caption[]{Temporal development of the difference between the number of turns (rotations)
of the Lorenz and sinusoidal oscillator, respectively, in the state of imperfect phase
synchronization. $\Omega=8100\,{\rm
rad/s}$, and crosses denote intersections with the Poincar\'e cross section $w=(R-1)/10$.
}
\label{tempdevel1}
\end{figure}

\begin{figure}
\includegraphics[width=3.2in]{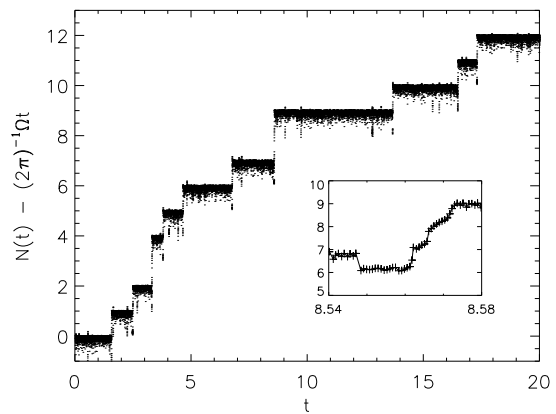}
\caption[]{Temporal development of the difference between the number of turns (rotations)
of the Lorenz and sinusoidal oscillator, respectively, in the state of imperfect phase
synchronization. $\Omega=8250\,{\rm
rad/s}$, and crosses denote intersections with the Poincar\'e cross section $w=(R-1)/10$.
}
\label{tempdevel2}
\end{figure}

Another demonstration of imperfect phase synchronization can be obtained
by plotting the temporal development between the phases of the driven Lorenz
system and the sinusoidal driving force (see Figures \ref{tempdevel1}--\ref{tempdevel2}).
As explained above, imperfect phase synchronization is characterized by the unbounded
character of return times, that leads to the driven system losing the pace of the
sinusoidal generator. Thus, at first sight it can be surprising ({\it e.g.}, from
Figure \ref{tempdevel2}) that the phase jumps ({\it i.e.}, errors of synchronization) are
quite often positive, as with the definition used this implies that the driven system actually 
performs more rotations than the sinusoidal driving (although in Figure \ref{tempdevel1}
one can find examples of both positive and negative jumps). The existence of these
jumps (far from the transition to nonsynchronization) is one of the well known
signatures of imperfect phase synchronization \cite{zaks99,park99}. It is also interesting
to mention that although jumps by one turn, $2\pi$ jumps in terms of phase, are the
most common, $4\pi$ can also be found (as in Fig.\ \ref{tempdevel2} for $t\in [8,9]$).

\begin{figure}
\includegraphics[width=3.2in]{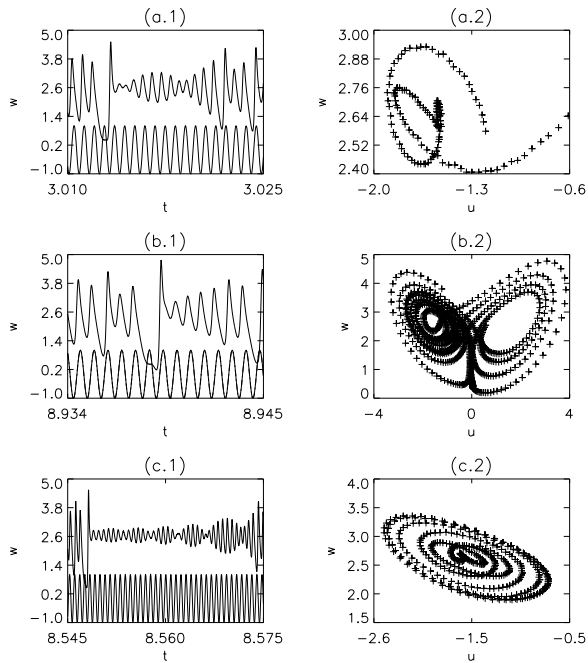}
\caption[]{Study of the behavior of the Lorenz system at three different phase jumps,
namely those represented in three insets of Figs. \ref{tempdevel1}-\ref{tempdevel2}.
The three left panels contain the evolution of variable $w$ for the three phase
jumps, respectively, while in the three right panels the phase portrait $w$ vs.\ $u$
is represented for a subset of the time interval.
Panel (b,2) represents the
whole time snapshot in panel (b,1), {\it i.e.}, $[8.934,8.945]$ while panels
(a,2) and (c,2) detail the fast turns happening in the time intervals 
$[3.0135,3.01525]$ for (a,1) and $[8.565,8.570]$ for (c,1), respectively.
}
\label{modul}
\end{figure}

The above mentioned {\it paradoxical\/} fact that typically the driven Lorenz system
performs more turns than the sinusoidal generator can be understood better by looking at
some time traces of one of the three state space variables, {\it e.g.} $w$, and also at
some state space projections (this is shown in Figure \ref{modul}). Considering variable
$w$ has the advantage that it can be compared more cleanly with the {\it sinusoidal
pacemaker\/} that is below in all the time traces (as $w$ remains always positive).
Anyhow, one has to keep in mind that the oscillations with a large period are 
associated with changing lobe (moment where the dynamics is more influenced by the 
saddle equilibrium). The results in Fig.\ \ref{modul}
correspond to three different phase jump events, that are the same presented in
the three insets of Figures \ref{tempdevel1}--\ref{tempdevel2}: a positive and a
negative, respectively, phase jump by $2\pi$ in Fig.\ \ref{tempdevel1} and a 
positive $4\pi$ jump in Fig.\ \ref{tempdevel2}. As explained above, the three phase 
jumps have in common that
they are preceded by a change of lobe in the Lorenz system (the slow turn in the left 
panels of Fig.\ \ref{modul}), and so the driven Lorenz oscillator loses almost one
turn when compared with the sinusoidal generator (this can also be seen clearly in the
three insets in Figures \ref{tempdevel1}--\ref{tempdevel2}).
Although the time from peak to peak (or, in other words, between two crossings
through the Poincar\'e plane) is not the same (it varies chaotically),
the variation happens in a relatively narrow range outside of these changes of lobe,
and the Lorenz system is able to keep the pace with the drive. However, when one of
these changes of lobe (and, thus, slow turns) occurs, 
the Lorenz system almost performs one turn less than the sinusoidal generator. These
events are relatively common as can be seen from the cloud of points going down below
the plateaus in Figures \ref{tempdevel1}--\ref{tempdevel2}, and that almost go down
to the level of one turn less (with respect to the level at the plateau).

However, the absence of phase jumps for many changes of lobe is due to the fact
that the Lorenz system is able to perform an extra rotation, with respect to the
sinusoidal generator.
When this does not happen, a negative phase jump occurs, while sometimes the Lorenz
system is, quite surprisingly, capable of performing two (or even three) extra
turns. These fast rotations (some of them could even be called {\it pseudo-rotations},
as they are characterized by a very small rotation radius, see panels (a,2) and (c,2)
in Fig.\ \ref{modul}) may happen inmediately after the change of lobe (panel (a,2)), 
or slightly after (panel (c,2)). Quite curiously, during these fast rotations the
variable $w$ exhibits an interesting modulational, or beating, transient periodic
behavior (resembling amplitude modulation). This behavior is probably
associated to the  interference between the {\it anomalous\/} fast rotations and the
frequency of the 
sinusoidal generator (of course, as negative phase jumps do not have associated fast 
rotations, the system does not exhibit modulational behavior in this instance).
These pseudo-rotations

On the other hand, if one looks carefully at the behavior of the system in the intervals
of time in which the system goes from an almost negative phase jump to a positive one 
(left inset of Fig.\ \ref{tempdevel1} and inset of Fig.\ \ref{tempdevel2}) one can 
see that the driven Lorenz system performs more turns than the sinusoidal drive. Following
Refs.\ \cite{zaks99,park99} one can interpret this behavior by a change in the locking ratio
between the drive and the Lorenz systems, that is no longer $1:1$, but rather $n:n+2$ (in a
$2\pi$ phase jump, or $n:n+3$ in a $4\pi$ phase jump (the locking ratio will be $n:n+1$
in the frequent events in which the system does not exhibit a phase jump, although there
is an almost negative phase jump, that occur at almost all changes of lobe). In this sense,
synchronization is not lost, but the system exhibits an alternation between different locking
rates, in the periods of type in which the dynamics is more strongly non-hyperbolic (those
in which the Lorenz system is under the effect of the saddle equilibrium point at the origin).

\section{Discussion} \label{secconc}

In the present study we have been able to characterize unambiguously imperfect
phase synchronization in a sinusoidally forced representation of the Lorenz oscillator
as an electronic (analog) circuit.
The results presented in this contribution are so clear and clean that sometimes are
almost identical to the equivalent results obtained from the direct numerical
simulation of the dynamical system (cf. Refs. \cite{zaks99,park99}), even though in 
our case the system is subject to sources of noise (thermal, channel, tolerances 
in the components, etc.) This precise correspondence between
experiment and numerical simulation makes us firmly believe 
that the imperfections observed in the phase synchronized state are not due to
the presence of noise, proximity to the onset of phase synchronization or the
like. In addition, the phase jumps have a clearly defined deterministic structure: 
during a transient period of time the system appears to be described by a different 
locking ratio (one would not expect this behavior in systems subject to noise or exhibiting
intermittent bursts). In addition, the close correspondence between theory and
experiment clearly confirm the reality of the phenomenon, and the possibility of
modeling it theoretically, as the obstructions to deterministic modelling discussed
in Refs.\ \cite{laigrekur99,laigreb99} do not apply to this case.

Chaotic (perfect) phase synchronization was first demonstrated from the analysis
of theoretical models \cite{rosenblum96}, and later has been demonstrated through
analog simulation of two coupled R\"ossler oscillators \cite{Parlitz96}, and
in some experimental physical systems: a plasma system \cite{Ticos00} and a chaotic
laser array \cite{deshazer01}. Imperfect phase synchronization
may be relatively common in dynamical systems with more degrees of freedom, and,
in fact, in Ref.\ \cite{park99} it was argued that it could be the mechanism
behind observations in some experimental data describing human cardiorespiratory
activity \cite{SchaeferNat,SchaeferPre}. 

The outlook of the present experimental demonstration is that imperfect phase
synchronization should be relatively common in a number of fields. The reason
for this is that unstable fixed points being part of the closure of a chaotic
attractor are relatively common in a number of fields, like fluid mechanics ({\it e.g.},
in the transition to turbulence), nonlinear optics ({\it e.g.}, semiconductor lasers), etc.
However, the behavior of the system can be more complex that the one presented here,
as the unstable fixed point at the origin of the Lorenz system is a saddle, while
higher-dimensional systems will typically have saddle-focus unstable fixed points.


\begin{acknowledgments}

We acknowledge financial support from MCyT (Spain) and FEDER through Grants BFM2000-1108 and
BFM2001-0341, and the European Commission through Projects OCCULT, IST-2000-29683 and
HPRN-CT-2000-00158.

\end{acknowledgments}



\begin{thebibliography}{99}

\bibitem{syncbook}
A. Pikovsky, M. Rosenblum, and J. Kurths, {\em Synchronization: A Universal
  Concept in Nonlinear Science} (Cambridge University Press, Cambridge, 2001).

\bibitem{ottbook}
E. Ott, {\em Chaos in Dynamical Systems} (Cambridge University Press,
  Cambridge, 1993).

\bibitem{fujisaka}
H. Fujisaka and T. Yamada, Prog. Theor. Phys. {\bf 69},  32  (1983).

\bibitem{pecora90}
L.~M. Pecora and T.~L. Carroll, Phys. Rev. Lett. {\bf 64},  821  (1990).

\bibitem{Rulkov95}
N.~F. Rulkov, M.~M. Sushchik, L.~S. Tsimring, and H.~D.~I. Abarbanel, Phys.
  Rev. E {\bf 51},  980  (1995).

\bibitem{rosenblum96}
M.~G. Rosenblum, A.~S. Pikovsky, and J. Kurths, Phys. Rev. Lett. {\bf 76},
  1804  (1996).

\bibitem{rossler}
O. R{\"o}ssler, Phys. Lett. A {\bf 57},  397  (1976).

\bibitem{yalcin97}
T. Yalcinkaya and Y.-C. Lai, Phys. Rev. Lett. {\bf 79},  3885  (1997).

\bibitem{pikovsky97a}
A.~S. Pikovsky, M.~G. Rosenblum, G.~V. Osipov, and J. Kurths, Physica D {\bf
  105},  219  (1997).

\bibitem{josicmar}
K. Josic and D.~J. Mar, Phys. Rev. E {\bf 64},  056234  (2001).

\bibitem{lorenz}
E.~N. Lorenz, J. Atmos. Sci. {\bf 20},  130  (1963).

\bibitem{zaks99}
M.~A. Zaks, E.~H. Park, M.~G. Rosenblum, and J. Kurths, Phys. Rev. Lett. {\bf
  82},  4228  (1999).

\bibitem{park99}
E.~H. Park, M.~A. Zaks, and J. Kurths, Phys. Rev. E {\bf 60},  6627  (1999).

\bibitem{zaks00}
M.~A. Zaks, E.~H. Park, and J. Kurths, Int. J. Bif. Chaos {\bf 10},  2649
  (2000).

\bibitem{Pikovsky97b}
A.~S. Pikovsky {\it et~al.}, Phys. Rev. Lett. {\bf 79},  47  (1997).

\bibitem{Pikovsky97d}
A. Pikovsky {\it et~al.}, Chaos {\bf 7},  680  (1997).

\bibitem{laigrekur99}
Y.~C. Lai, C. Grebogi, and J. Kurths, Phys. Rev. E {\bf 59},  2907  (1999).

\bibitem{laigreb99}
Y.~C. Lai and C. Grebogi, Phys. Rev. Lett. {\bf 82},  4803  (1999).

\bibitem{sanchezphd}
E. S\'anchez, Ph{D} {T}hesis, Universidad de Salamanca, Spain, 1999.

\bibitem{sanchezpre98}
E. S\'anchez and M.~A. Mat\'{\i}as, Phys. Rev. E {\bf 57},  6184  (1998).

\bibitem{Parlitz96}
U. Parlitz, L. Junge, W. Lauterborn, and L. Kocarev, Phys. Rev. Lett. {\bf 54},
   2115  (1996).

\bibitem{Ticos00}
C.~M. Ticos {\it et~al.}, Phys. Rev. Lett. {\bf 85},  2929  (2000).

\bibitem{deshazer01}
D.~J. DeShazer, R. Breban, E. Ott, and R. Roy, Phys. Rev. Lett. {\bf 87},
  044101  (2001).

\bibitem{SchaeferNat}
C. Sch{\"a}fer, M.~G. Rosenblum, H.-H. Abel, and J. Kurths, Nature (London)
  {\bf 392},  239  (1998).

\bibitem{SchaeferPre}
C. Sch{\"a}fer, M.~G. Rosenblum, H.-H. Abel, and J. Kurths, Phys. Rev. E {\bf
  60},  857  (1999).

\end{thebibliography}
\end{document}